\documentclass[final,3p,twocolumn]{elsarticle}

\makeatletter
\def\ps@pprintTitle{%
  \let\@oddhead\@empty
  \let\@evenhead\@empty
  \let\@oddfoot\@empty
  \let\@evenfoot\@oddfoot\
}
\makeatother
\usepackage{tabularx}
\usepackage[pagewise]{lineno}
\modulolinenumbers[1]
\usepackage{hyperref}
\usepackage{array}
\usepackage{svg}
\hypersetup{colorlinks=true,pdfborder=0 0 0}
\usepackage{subcaption}
\usepackage{amsmath,bm,siunitx,amssymb,caption,booktabs}
\usepackage{graphicx}
\usepackage{multirow}

\modulolinenumbers[5]
\usepackage{textcomp,gensymb}
\usepackage{cleveref}
\usepackage{ulem}
\usepackage{xcolor}

\usepackage{fancyhdr}

\usepackage{etoolbox}
\makeatletter
\patchcmd{\ps@pprintTitle}% <cmd>
  {Preprint submitted}% <search>
  {Postprint of article submitted}% <replace>
  {}{}% <succes><failure>
\makeatother

\journal{Acta Materialia}

\begin{document}

\begin{frontmatter}

\title{Eutectic and peritectic equilibria in \\ coherent binary alloys}

\address[az-mse]{Department of Materials Science and Engineering, University of Arizona, Tucson, AZ 85721, USA}
\address[az-am]{Graduate Interdisciplinary Program in Applied Mathematics, University of Arizona, Tucson, AZ 85721, USA}

\author[az-mse]{Samiah Hassan}
\author[az-mse]{Jiayang Wang}
\author[az-am]{Teddy Meissner}
\author[az-mse]{Pierre A. Deymier}
\author[az-mse,az-am]{Marat I. Latypov\corref{cor1}}
\cortext[cor1]{corresponding author}
\ead{latmarat@arizona.edu}

\begin{abstract}

This work extends the Cahn--Larché thermodynamic framework to binary alloys in which two coherent solid phases coexist with an incoherent liquid and investigates how coherency strain energy modifies classical eutectic and peritectic equilibria. We derive equilibrium conditions for three-phase coexistence that include an elastic energy term dependent on the molar fractions of the solid phases and apply them to model binary eutectic and peritectic systems. We find that coherency stress transforms the eutectic point into a finite three-phase equilibrium field spanning a continuous range of compositions and temperatures. In systems whose equilibrium compositions are peritectically ordered, coherency stress progressively destabilizes the two-solid equilibrium without generating a stable three-phase field and can suppress the peritectic reaction entirely. This asymmetry is governed by the geometric relationship between the compositions of the phases: when the liquid composition lies between those of the two solids (eutectic configuration), the liquid serves as a thermodynamic buffer against the coherency penalty on the solid--solid pair; when it lies outside (peritectic configuration), no such mechanism is available. These results demonstrate that coherency stress can fundamentally alter three-phase  equilibria involving a liquid and suggest that such effects may be significant in systems with large coherent misfits. \\

\noindent Note: \textcolor{red}{this is an author-generated postprint} of the article by Hassan et al.\ \href{https://doi.org/10.1016/j.actamat.2026.122841}{published} in {\it{Acta Mater.}}. \\ DOI for version of record: 10.1016/j.actamat.2026.122841

\end{abstract}


\end{frontmatter}

\pagestyle{fancy}
\fancyhf{}
\fancyhead[LO]{Postprint of \href{https://doi.org/10.1016/j.actamat.2026.122841}{Hassan et al., Acta Mater (2026) 122841}}

% \linenumbers

\section{Introduction} 
\label{sec:intro}

Coherency stress is a common feature of multiphase crystalline solids in which continuity of the lattice across an interface accommodates a mismatch between the stress-free lattices of the phases \cite{porter2009phase}. Coherent interfaces are commonly found in technologically critical materials. Examples include Ni-base superalloys with the $\gamma/\gamma^\prime$ microstructure providing high-temperature strength \cite{pollock2006nickel}; age-hardened aluminum alloys and advanced steels strengthened with coherent precipitates \cite{porter2009phase}; and strain-engineered semiconductor heterostructures that exploit coherent epitaxy to tailor electronic band structure in devices based on Si--Ge, Ge--Sn, and III–V alloys \cite{zunger1989structural,stringfellow2021epitaxial,JUNG2025121369}. Coherent microstructures are increasingly relevant with the rise of processing and manufacturing methods that involve rapid solidification (e.g., metal 3D printing), which can preserve coherent interfaces between phases that would otherwise relax through misfit-dislocation formation under slower cooling \cite{debroy2018additive}.

The strain energy associated with an elastically accommodated lattice misfit  between coherent phases contributes to the total free energy of the system and fundamentally alters the conditions for phase equilibria. The general framework for treating thermodynamics of stressed solids was established by Larché and Cahn in a series of foundational studies \cite{larche1973linear,larche1978nonlinear,larche1985overview}. Applying this framework to a model binary system composed of two isotropic, elastically identical solid phases, Cahn and Larché \cite{cahn1984simple} showed that coherency strains stabilize a single-phase solid solution relative to a two-phase mixture, that the classical common-tangent construction does not apply, and that the Gibbs phase rule no longer holds: the number of coexisting phases becomes independent of the classical thermodynamic degrees of freedom. Subsequent theoretical work by Johnson and Voorhees \cite{johnson1987phase,johnson1988coherent} further elucidated the consequences of coherency stress, including the existence of multiple linearly stable equilibrium states under identical thermodynamic conditions. The predictions of the Larché--Cahn theory have received direct experimental confirmation by Shi et al.\ \cite{shi2018verifying}, who verified the open-system elastic parameters in nanoporous Pd–H using dynamic mechanical analysis and found excellent agreement with theory. More recent generalizations have extended the Cahn–Larché model to systems with free surfaces \cite{spatschek2016scale}, large lattice mismatch \cite{phan2019coherent}, and elastic anisotropy \cite{phan2019modeling}.

Beyond phenomenological formulations based on the Cahn–Larché theory, coherent phase diagrams have been pursued with computational methods. Wood and Zunger \cite{wood1988epitaxial,WoodZunger1989} introduced a first-principles approach using density functional theory and the cluster variation method to calculate coherent phase diagrams. More recently, Behara et al.\ \cite{behara2024chemomechanics} developed a statistical-mechanics framework coupling chemical ordering and mechanical strain from first principles to predict alloy phase stability under coherent conditions. Using the same approach, Jung et al.\ \cite{JUNG2025121369} examined how strain and composition affect the thermodynamic stability of Ge--Sn alloys and demonstrated new design spaces for strain-engineered semiconductor materials. Wang and Chen \cite{wang2024theory} developed a general thermodynamic theory for strain phase equilibria and diagrams applicable to diverse material systems including ferroelectrics and correlated oxides. Coherency stress effects have also been of continuing interest in the phase-field community, from the classical works of Khachaturyan with coworkers \cite{khachaturyan2008theory,wang1995microstructurala,wang1995microstructuralb} to recent chemomechanical phase-field developments \cite{shen2009phase,rudraraju2016mechanochemical}.

Despite this extensive body of work, nearly all studies of coherency effects on phase equilibria have focused on fully solid-state systems. Although a liquid cannot support shear stress, coherency between two coexisting solids can still contribute to the total free energy and thus change conditions for phase equilibria. Among the few studies to address this, Roytburd and Sung \cite{roytburd1993formation} showed that elastic interactions in a thin-film eutectic system grown epitaxially on a substrate alter the topology of the eutectic phase diagram by splitting the eutectic point into two points or transforming it into a consolute point, depending on the sign of the elastic interaction energy between the phases. In all cases, three-phase coexistence in their model remains confined to isolated points in composition–temperature space. Our own prior work examined the effect of epitaxial strain on solid–liquid equilibria involving a chemical reaction \cite{deymier2023thermodynamics} and on dissolution/crystallization of apatite under stress \cite{deymier2023effect}. 
However, the effect of intrinsic coherency stress from lattice misfit between two solid phases on the classical invariant equilibria has not been investigated.

This gap is critical because eutectic and peritectic reactions govern solidification phenomena, including freezing ranges, solute partitioning between solid and liquid, as well as microstructural length scales. They also control the spatial extent of the mushy zone, which is the semi-solid region where liquid and solid coexist during casting, welding, and additive manufacturing \cite{dantzig2016solidification,flemings1974solidification}. Classical models of eutectic and peritectic solidification rely on conventional stress-free phase diagrams \cite{jackson1988lamellar,boettinger2000solidification,worster1997convection} and thus overlook potential coherency effects.

To examine these effects, we present a phenomenological extension of the Cahn--Larch\'e two-solid model to systems containing an incoherent liquid. We derive equilibrium conditions within this model (\Cref{sec:model}). Model eutectic and peritectic systems (introduced in \Cref{sec:gibbs}) exhibit different responses to the coherency effects (\Cref{sec:results}). \Cref{sec:discussion} formalizes this observation and discusses the magnitude of the coherency effects in the context of elastic properties of real materials.

\section{Theoretical model} 
\label{sec:model}

Our study considers binary alloy systems consisting of two elements $A$ and $B$. These systems form solid phases at low temperatures: $A$-rich $\alpha$ phase and $B$-rich $\beta$ phase with limited solubility of the other element. The systems can form an invariant point in which both solid phases are in equilibrium with a liquid phase, $L$. Our goal is to construct a theoretical model that provides conditions for such three-phase equilibria in the presence of a coherency stress between the solid phases. To facilitate analytical solution, we use simplifying assumptions about the system similar to those used in the original Cahn--Larché model for two-phase systems \cite{cahn1984simple}. Specifically, we consider an infinite volume occupied by three phases; the solid phases are isotropic and linearly elastic; both solid phases have the same elastic properties described by, e.g., Young’s modulus $E$ and Poisson’s ratio $\nu$. We further assume that the solid--liquid interface is incoherent so that the liquid phase does not support coherency stress. 
With these assumptions, we consider the coherency contribution per mole of the entire system to be proportional to the solid molar fractions, $z_\alpha$ and $z_\beta$:

\begin{equation}
g_e = z_{\alpha}z_{\beta} \frac{VE\epsilon^2}{(1-\nu)} = z_\alpha z_\beta \mathcal{W},
\label{eq:g_e}
\end{equation}

\noindent where we introduce the coherency energy parameter $\mathcal{W} = VE\epsilon^2/(1-\nu)$. Here $V$ is a common reference molar volume of the solids (\si{\cubic\metre\per\mole}) and $\epsilon$ is the lattice misfit between the stress-free $\alpha$ and $\beta$ lattices. In the absence of liquid ($z_L=0$), \Cref{eq:g_e} follows from the coherent strain energy density $Y\eta^2(x-X)^2$ introduced by Cahn for an elastically isotropic solid, with $Y = E/(1-\nu)$ and $\eta$ the compositional expansion coefficient \cite{cahn1961spinodal,cahn1962spinodal}. We consider $\mathcal{W}$ independent of composition, phase fractions, and temperature in this study and use kJ/mol as its unit ($E$ in Pa corresponds to $\mathcal{W}$ and $g_e$ in J/mol). The coherency energy contribution, $g_e$, is additive to the chemical energies of the phases represented by molar Gibbs free energies, $g_i$ (with $i\in  \left\{ \alpha,\beta,L \right\} $), that constitute the total molar energy of the system, $g$: 

% \begin{equation}
% g(z_{\alpha}, z_{\beta}, x_{\alpha}, x_{\beta}, x_{L}) = z_{L}g_{L}(x_{L}) + z_{\alpha}g_{\alpha}(x_{\alpha}) + z_{\beta}g_{\beta}(x_{\beta}) + z_\alpha z_\beta \mathcal{W},
% \label{eq:g_total}
% \end{equation}

\begin{equation}
\begin{split}
g(z_{\alpha}, z_{\beta}, x_{\alpha}, x_{\beta}, x_{L}) = \\
z_{L}g_{L}(x_{L}) + z_{\alpha}g_{\alpha}(x_{\alpha}) + z_{\beta}g_{\beta}(x_{\beta}) + z_\alpha z_\beta \mathcal{W},%g_e(z_{\alpha},z_{\beta},\mathcal{W}), 
\end{split}
\label{eq:g_total}
\end{equation}

\noindent where $x_i$ is the composition of the $i$th phase expressed as a concentration of element $B$: $x_i$ = $N^B_i/(N^A_i+N^B_i)$ with $N^A_i$ and $N^B_i$ representing the number of moles of $A$ and $B$ elements in the $i$th phase. For an alloy of the overall composition, $X$, the phase molar fractions and compositions must satisfy the mass balance and the ranges given their definitions: 

\begin{subequations}
\label{eq:constr}
\begin{align}
X = (1-z_{\alpha}-z_{\beta})x_{L}+z_{\alpha}x_{\alpha}+z_{\beta}x_{\beta}, \\
z_{\alpha}+z_{\beta}+z_{L}=1, \\
0\leq z_{\alpha},z_{\beta},z_{L},x_{\alpha}, x_{\beta}, x_{L},X\leq 1. 
\end{align}

\end{subequations}

\noindent Here, $X$ is also expressed as a total molar concentration of $B$: $X = N^B/(N^A+N^B)$, $N^A$ and $N^B$ denoting the total moles of $A$ and $B$ elements in the alloy. 

Minimizing the total Gibbs free energy in \Cref{eq:g_total} subject to the constraints in \Cref{eq:constr} leads to the conditions for equilibrium in terms of $z_i$ and $x_i$ values. Candidates for the global minimum (stationary conditions) can be found using the method of Lagrange multipliers. To this end, we introduce a Lagrangian: 

% \begin{equation}
% \begin{split}
% \mathcal{L} = z_L g_L(x_L) + z_{\alpha} g_\alpha(x_{\alpha}) + z_{\beta} g_\beta(x_{\beta}) + \mathcal{W} z_{\alpha} z_{\beta} \\- \lambda (z_L x_L + z_{\alpha} x_{\alpha} + z_{\beta} x_{\beta} - X),
% \label{eq:lagrange}
% \end{split}
% \end{equation}

\begin{equation}
\begin{split}
% \mathcal{L}(z_{\alpha}, z_{\beta}, z_L, x_{\alpha}, x_{\beta}, x_L, \lambda)
\mathcal{L} = z_L g_L(x_L) + z_{\alpha} g_\alpha(x_{\alpha}) + z_{\beta} g_\beta(x_{\beta}) + \\
+ \mathcal{W} z_{\alpha} z_{\beta} - \lambda (z_L x_L + z_{\alpha} x_{\alpha} + z_{\beta} x_{\beta} - X),
\label{eq:lagrange}
\end{split}
\end{equation}

\noindent where $\lambda$ is a Lagrange multiplier and the molar fraction of liquid is not independent $z_L = (1-z_\alpha-z_\beta)$. Requiring partial derivatives of the Lagrangian with respect to the six independent variables ($z_\alpha$, $z_\beta$, $x_\alpha$, $x_\beta$, $x_L$, $\lambda$) to be equal to zero gives candidate three-phase stationary states when all phase fractions are positive and all compositions are inside their allowed interval.  The resulting six equations for the partials can be arranged as follows: 

% \begin{equation}
% \frac{dg_L}{dx_{L}} = \frac{dg_\alpha}{dx_{\alpha}} = \frac{dg_\beta}{dx_{\beta}}=\frac{g_\alpha(x_{\alpha}) - g_L(x_{L}) + {\mathcal{W}} z_\beta}{x_{\alpha} - x_{L}} = \frac{g_\beta(x_{\beta}) - g_L(x_{L}) + {\mathcal{W}} z_\alpha}{x_{\beta} - x_{L}}.
% \label{eq:eqs}
% \end{equation}

\begin{equation}
\begin{split}
\frac{dg_L}{dx_{L}} &= \frac{dg_\alpha}{dx_{\alpha}} = \frac{dg_\beta}{dx_{\beta}} = \\
&=\frac{g_\alpha(x_{\alpha}) - g_L(x_{L}) + {\mathcal{W}} z_\beta}{x_{\alpha} - x_{L}} = \\
&= \frac{g_\beta(x_{\beta}) - g_L(x_{L}) + {\mathcal{W}} z_\alpha}{x_{\beta} - x_{L}}.
\end{split}
\label{eq:eqs}
\end{equation}

Solving these equations for a specific set of functions describing the Gibbs free energies of individual phases (e.g., from thermodynamic databases) provides candidate conditions for the thermodynamic equilibrium of the three phases as they can represent not only minima but also saddle points. 

When the coherency term is zero, $\mathcal{W}=0$, \Cref{eq:eqs} recovers classical equations that graphically manifest in the well-known common-tangent construction. The classical common tangent also follows from \Cref{eq:eqs} for solid--liquid pairs because the elastic term vanishes when $z_\alpha=0$ or $z_\beta=0$. With the non-zero coherency stress, $\mathcal{W}\ne0$, and in the presence of both solid phases, the common tangent construction does not hold.  For the two-solid states in the systems considered below, the coherent energy lies above the stress-free common tangent and curves downward with a curvature that depends on the coherency magnitude (see \Cref{fig:eu_gibbs,fig:P_gibbs}). Cahn and Larché's problem of two coherent solids can be recovered from \Cref{eq:g_total} at $z_L = 0$, i.e., in the no-liquid limit of our framework.

Solving the equations (\Cref{eq:eqs}) with $\mathcal{W}\ne0$ for a range of alloy compositions and temperatures, $(X,T)$ allows us to obtain phase equilibria in maps reminiscent of phase diagrams. We refer to them as {\it{phase equilibria maps}} because, unlike phase diagrams, phase compositions here depend on the overall alloy composition. Further, construction using fixed tie-line endpoints does not apply, however, phase equilibria maps still satisfy mass balance, including the lever rule using the actual compositions of two coexisting phases (from curved tangents in case of coherent solids). \Cref{sec:results} presents implementation and results of this approach for specific eutectic- and peritectic-forming binary systems.
 
\section{Free-energy parametrization and numerical implementation}
\label{sec:gibbs}

\subsection{Functional forms of the Gibbs free energies}
\label{sec:gforms}

We present two case studies that use the general theoretical model presented in the previous section to elucidate the impact of coherency stress on invariant points. The two case studies consider (i) a eutectic-forming binary alloy system and (ii) a binary system with a peritectic. To focus on the fundamental and qualitative effects of the coherency stress, we consider simple quadratic functions to describe the molar Gibbs free energies of the solid phases ($\alpha$ and $\beta$) parametrized by unique sets of scalar parameters, $x^0$, $a$, $b$: 

\begin{subequations}
\label{eq:gs}
\begin{align}
g_\alpha = a(x_{\alpha} - x^0_{\alpha})^2 + b_\alpha \\
g_\beta = a(x_{\beta} - x^0_{\beta})^2 + b_\beta
\end{align}
\end{subequations}

\noindent We use a similar functional form for the liquid phase, $L$ with the addition of a linear dependence on  
temperature, $T$: 
\begin{equation}
g_L = a(x_{L} - x^0_{L})^2 + b_L - {\varphi}T, 
\label{eq:gL}
\end{equation}

\noindent where the $\varphi$ coefficient controls the rate at which the liquid phase is stabilized with increasing temperature and thus implicitly encodes the entropy of fusion, $\varphi = \Delta S_f$, and is discussed in this sense in \Cref{sec:discussion}. In both \Cref{eq:gs,eq:gL}, the $b_i$ parameter sets the reference free energy of each phase at its reference composition $x^0_i$, which corresponds to the minimum of $g_i$.

The adopted quadratic form is the second-order Taylor expansion of a smooth molar Gibbs energy about its minimum, and is therefore the minimal parametrization that preserves the key features pertinent to the present analysis: a well-defined reference composition for each phase, $x^0_{i}$, and a finite curvature $\mathrm{d}^2 g_i/\mathrm{d}x_i^2 = 2a_i$. Parabolic free energies of this form are common in phase-field modeling and its coupling to thermodynamic databases. For example, parabolic functions fitted to the CALPHAD assessment of two binary aluminum-base systems were reported to reproduce the same common tangents as the full functions \cite{hu2007thermodynamic,welland2017method}. Cahn and Larch\'e \cite{cahn1984simple} likewise used a minimal chemical free energy in order to isolate the elastic contribution.

In this study, the quadratic parametrization of the Gibbs free energy allows us to investigate both eutectic and peritectic-forming binary systems by a suitable selection of the $x^0$ parameters that control the position of the parabolic curves. A eutectic point can be introduced by setting $x^0_\alpha < x^0_L < x^0_\beta$, whereas $x^0_L < x^0_\alpha < x^0_\beta$ represents a system with a peritectic point. Although simple, this parametrization allows us to elucidate fundamental effects of coherency stress on three-phase equilibria at invariant points. To this end, we substitute these Gibbs free energy functions (\Cref{eq:gs,eq:gL}) into \Cref{eq:eqs} and solve the resulting system of equations for a given combination of alloy composition, $X$, temperature, $T$, and coherency term, $\mathcal{W}$. Solving the equations provides us with molar fractions, $z_i$ and concentrations $x_i$ (e.g., of element $B$) of the phases in equilibrium under given conditions. Obtaining phase molar fractions for systematically varied alloy compositions, $X$, and temperatures, $T$, allow us to construct phase equilibria maps to visualize the impact of coherency stress on eutectic and peritectic regions of binary phase diagrams. 

\begin{table*}[h!]
\centering
\caption{Parameters adopted for the Gibbs energy functions of the two systems. The reference compositions $x^0_\alpha$, $x^0_\beta$ and $x^0_L$ are dimensionless mole fractions; the curvature $a$ and the reference energies $b_\alpha$, $b_\beta$ and $b_L$ are in \si{\kilo\joule\per\mole}; the temperature coefficient $\varphi$ is in \si{\kilo\joule\per\mole\per\kelvin}.}
\begin{tabular}{lc*{6}{>{\centering\arraybackslash}m{3.5mm}}c}
\toprule
\textbf{Parameter}
& $x_\alpha^{0}$ & $x_\beta^{0}$ & $x_L^{0}$ & $a$
& $b_\alpha$ & $b_\beta$ & $b_L$ & $\varphi$ \\
\hline
\textbf{Eutectic system}
& 0.2 & 0.8 & 0.5 & 50 & 21 & 21 & 40 & 0.0167 \\
\hline
\textbf{Peritectic system}
& 0.5 & 0.8 & 0.3 & 50 & 21 & 21 & 40 & 0.0167 \\
% \hline
\bottomrule
\end{tabular}
\label{tab:parameters}
\end{table*}

\subsection{Parameter values for the model eutectic and peritectic systems}
\label{sec:param}

For the functional forms in \Cref{eq:gs,eq:gL}, we first adopt "symmetric" parameter sets that simplify the analysis and then test the robustness of the findings to asymmetric parametrizations within the quadratic representation (\Cref{sec:sensitivity}).

The symmetric parametrization assigns the same curvature, $a$, for all three free energy functions and the same reference energy, $b$, for the two solid phases: $b_\alpha = b_\beta = b < b_L$. Using the identical curvature, $a$, for all phases simplifies the analysis because it makes the temperature extent of the three-phase equilibrium independent of the curvature (\Cref{eq:dT}). \Cref{tab:parameters} lists the numerical values adopted for the two systems within the symmetric parameterization. This parameters selection describes binary systems where the melting points of the solid phases are comparable and are both much higher than the temperature of the invariant point.  The invariant point is set by the parameters as $T_\text{inv} = (b_L - b_\alpha)/\varphi$. Guided by this relationship, we adopt $\varphi = \SI{16.7}{\joule\per\mole\per\kelvin}$ so that it results in a realistic invariant temperature: $T_\text{inv}=$ \SI{1137.7}{\kelvin}.

We finally note that only differences in the reference energies, $b_i$, are physically meaningful as adding a constant to all three parameters does not change the equilibrium conditions. The relevant chemical scale in this study is therefore $b_L - b = \SI{19}{\kilo\joule\per\mole}$, which plays the role of the enthalpy of fusion. Since this scale is very close to $b=\SI{21}{\kilo\joule\per\mole}$, we express $\mathcal{W}$ in units of $b$ in subsequent analysis as a concise and dimensionless measure of the elastic strain energy. The two nonzero values ($\mathcal{W}=1$ and \SI{3}{\kilo\joule\per\mole}) adopted for analysis in \Cref{sec:results} then become $\mathcal{W}\approx0.05b$ and $0.15b$.

\subsection{Numerical implementation and verification}
\label{sec:comp}

To streamline the solution of the systems of equations for many combinations of ($X$, $T$) required for the construction of phase equilibria maps, we developed the following algorithm with a Python implementation. 

\begin{enumerate}

    \item Solve equations for pairwise two-phase equilibria in solid phases and the liquid ($\alpha$--$L$, $\beta$--$L$).
    \item Solve equations for two-phase equilibria including the two solid phases ($\alpha$--$\beta$).
    \item Solve equations for three-phase equilibria including all three phases ($\alpha$--$\beta$--$L$). 
    \item Calculate the Gibbs free energies corresponding to seven potential cases of equilibrium: three single-phase, three two-phase, and one three-phase conditions. 
    \item For each overall alloy composition, $X$, and temperature, $T$, find the equilibrium that corresponds to the lowest Gibbs free energy and store the corresponding label that lists phases with non-zero molar fractions. 
    \item Visualize the results as fields of labels that list the phases in equilibrium. 
    
\end{enumerate}

Our implementation uses the library \texttt{sympy} \cite{meurer2017sympy} for symbolic solution of systems of equations in Python. 
For the quadratic free energies of \Cref{sec:gforms}, the stationary conditions for every combination of phases can also be reduced to a single polynomial equation  in terms of the diffusion potential (Appendix~\ref{app:mu}). An implementation of this reduced formulation serves as an independent verification of the symbolic solution by direct numerical minimization of \Cref{eq:g_total} over the phase fractions (Appendix~\ref{app:reduced}) without \Cref{eq:eqs}. Both implementations select the same equilibrium at every $(X,T)$ grid point of the phase equilibria maps shown in the next section. The reduced formulation is further used for systems with asymmetric parameter sets analyzed in \Cref{sec:sensitivity}. 

\section{Results}
\label{sec:results}

\subsection{Case study 1: Eutectic system}

\begin{figure*}[h]
  \centering
  \includegraphics[width=\linewidth]{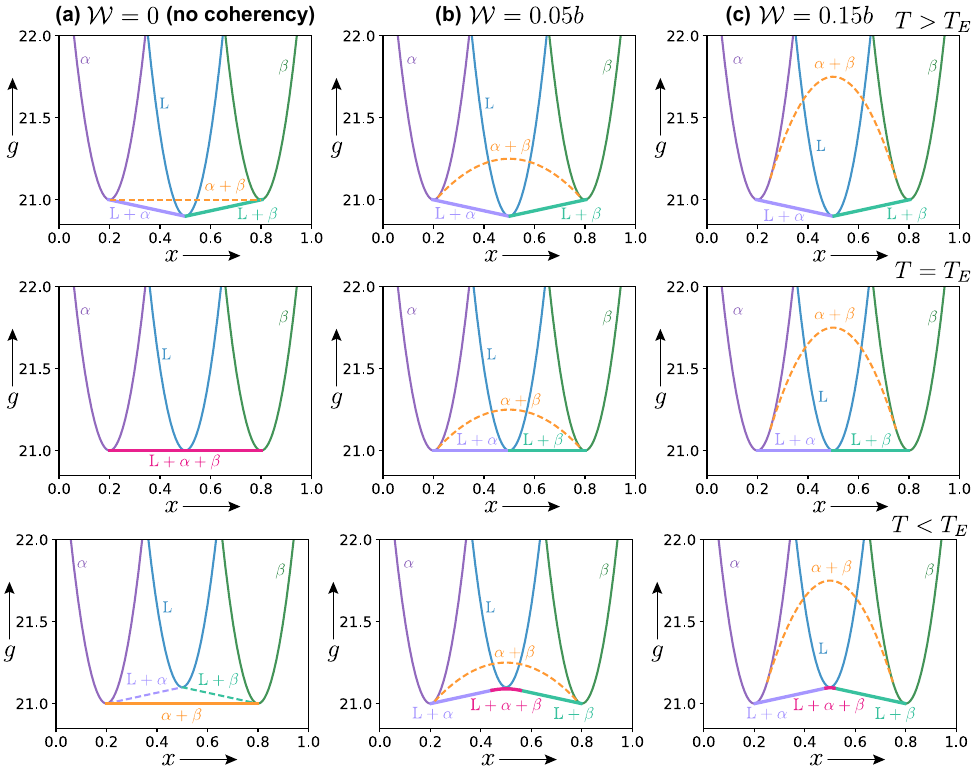}
  \caption{Gibbs free energy curves at three representative temperatures relative to the eutectic temperature in the stress-free system ($T_E$) obtained with various coherency magnitudes, $\mathcal{W}$, expressed in terms of the reference free energy parameter of the solid phases ($b=b_\alpha=b_\beta$). Solid curves represent free energies of single phases (purple $\alpha$, green $\beta$, blue $\text{L}$); thicker and dashed solid lines represent two-phase common tangents (orange $\alpha+\beta$, purple $\text{L}+\alpha$, green $\text{L}+\beta$), coherent $\alpha+\beta$ are represented by orange dashed curves, three-phase equilibria $\text{L}+\alpha+\beta$ are in solid magenta (line in (a), curves in (b,c)).}
  \label{fig:eu_gibbs}
\end{figure*}

\begin{figure*}[h]
  \centering
  \includegraphics[width=\linewidth]{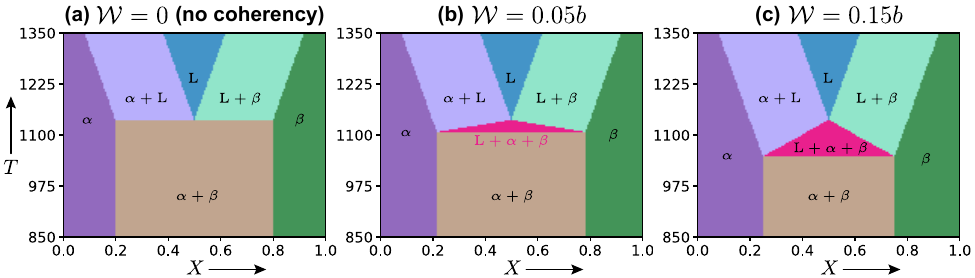}
  \caption{Phase equilibria maps for various coherency magnitudes, $\mathcal{W}$, expressed in terms of the reference free energy parameter of the solid phases ($b=b_\alpha=b_\beta$).}
  \label{fig:eutectic-pd}
\end{figure*}

We first consider the application of our theoretical model (\Cref{sec:model}) for a binary system that includes a eutectic point in its conventional stress-free phase diagram. To this end, we set $x^0_\alpha < x^0_L < x^0_\beta$ in the Gibbs free energy parametrization (\Cref{eq:gs,eq:gL}). We then analyze the Gibbs free energy curves corresponding to different combinations of the phases (\Cref{fig:eu_gibbs}) and phase equilibria maps (\Cref{fig:eutectic-pd})  for a full range of overall alloy compositions, $X\in[0,1]$, and a temperature range, \SIrange{850}{1350}{\kelvin}, which covers the conditions for three-phase eutectic equilibria in this system.

 The Gibbs free energy curves clearly show the impact of coherency strain effects on the energetics of the system (\Cref{fig:eu_gibbs}). The Gibbs free energy curves are shown at three magnitudes of the $\mathcal{W}$ parameter and at three temperatures: (i) below, (ii) equal to, and (iii) above the eutectic temperature, $T_E$, in the stress-free phase diagram. In the absence of the coherency stress ($\mathcal{W}=0$, \Cref{fig:eu_gibbs}a), our calculations reproduce the traditional common-tangent construction. The lowest energy configurations follow the standard sequence: $\alpha$ and $\beta$ at $T=T_1 < T_E$, three-phase equilibrium at $T=T_2 = T_E$, and pairwise liquid--solid equilibria at $T=T_3 > T_E$. 

The addition of the coherency stress ($\mathcal{W}>0$, \Cref{fig:eu_gibbs}(b,c)) leads to distinct changes. First, the curve representing equilibrium between the solid phases arches upward instead of the classical linear common tangent. Consequently, the liquid phase becomes energetically favorable at temperatures where it would be unstable in a classical, incoherent system. For example, the liquid curve lies below the coherent two-solid curve even at $T \geq T_E$ (first two rows in \Cref{fig:eu_gibbs}(b,c)). Three-phase equilibria are also observed as the lowest energy configuration at a temperature below $T_E$ (see red segment in \Cref{fig:eu_gibbs}(b,c)) and, even more interestingly, for a range of compositions. Finally, this compositional range of three-phase equilibria is narrower for a greater magnitude of the $\mathcal{W}$ parameter: compare the red segments in \Cref{fig:eu_gibbs}(b) vs.\ \Cref{fig:eu_gibbs}(c). 

By systematically solving the equilibrium equations over the full range of temperatures and compositions, we translate the energetic shifts observed in the \Cref{fig:eu_gibbs} into the phase equilibria maps presented in \Cref{fig:eutectic-pd}. These maps visualize the stability domains of the liquid ($L$), solid ($\alpha$, $\beta$), and mixed-phase regions as a function of the coherency term, $\mathcal{W}$. In the stress-free limit ($\mathcal{W}=0$, \Cref{fig:eutectic-pd}(a)), our calculated map reproduces a classical binary eutectic phase diagram. The three phases ($\alpha$, $\beta$, and $L$) coexist only at a single eutectic temperature consistent with the classical Gibbs phase rule. 

Introducing coherency stress fundamentally alters the topology of the phase equilibria maps. As $\mathcal{W}$ increases, the eutectic equilibrium expands into a distinct three-phase field ($\alpha$+$\beta$+L). This region of three-phase coexistence grows in height with increasing coherency term but narrows in width, consistent with the curves in \Cref{fig:eu_gibbs}. These trends can be quantified by expressing both the temperature extent, $\Delta T$, and the compositional width, $\Delta X$, of the three-phase field as a function of $\mathcal{W}$ (Appendices~\ref{app:field} and~\ref{app:width}). 

The liquidus boundaries between $L$ and a solid--liquid region remain unchanged because $g_e=0$ when either solid is absent. The lower boundary of liquid-containing equilibrium is depressed by the penalty on two-solid coexistence. For the symmetric eutectic parameters in \Cref{tab:parameters}, the three-phase field extends from $T_E-\mathcal{W}/(2\varphi)$ to $T_E$, giving temperature extents of 29.9 and \SI{89.8}{\kelvin} at the two nonzero $\mathcal{W}$ values. The upper vertex of the three-phase field is pinned at the stress-free eutectic temperature, $T_E$, for all $\mathcal{W}$ (at which the field exists) because the corresponding condition is independent of $\mathcal{W}$ (Appendix~\ref{app:field}).

\subsection{Case study 2: Peritectic system}

We next use our model for the binary system that includes a peritectic point in its conventional stress-free phase diagram. To this end, we set $x^0_L < x^0_\alpha < x^0_\beta$ in the functions describing the Gibbs free energies of the phases (\Cref{eq:gs,eq:gL}). As for the eutectic case, \Cref{tab:parameters} lists a specific set of values adopted for the parameters defining the $g$ functions. This parametrization ensures a peritectic point within the considered temperature range (\SIrange{850}{1350}{\kelvin}). 

\begin{figure*}[h]
  \centering
  \includegraphics[width=\linewidth]{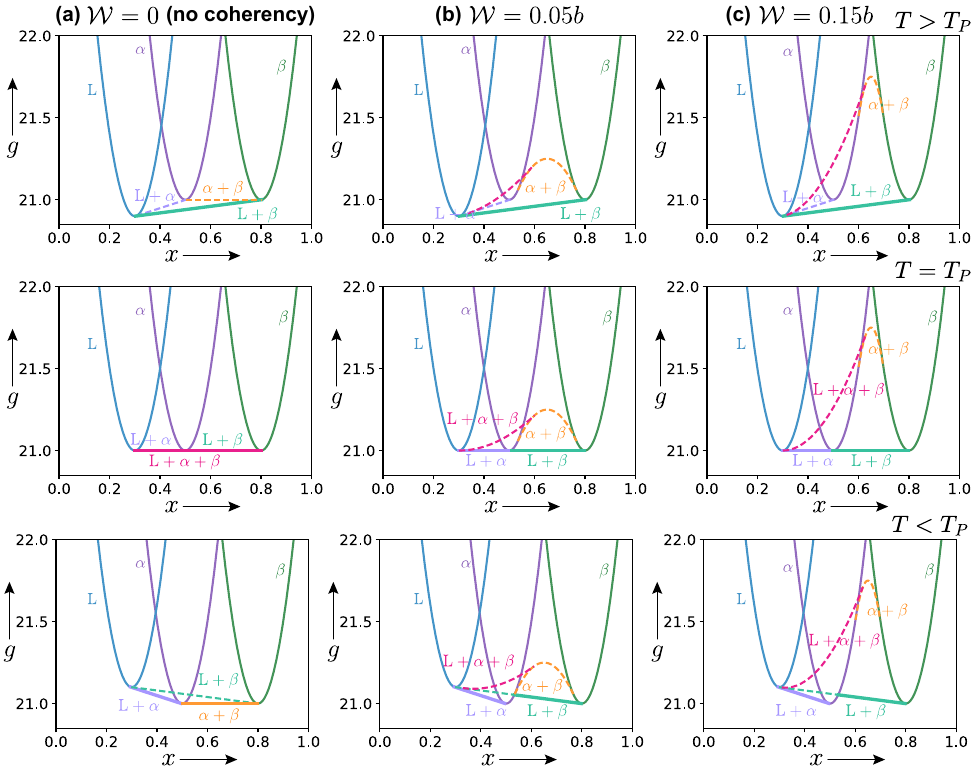}
  \caption{Gibbs free energy curves at three representative temperatures relative to the peritectic temperature in the stress-free system ($T_P$) obtained with various coherency magnitudes, $\mathcal{W}$. Color code is the same as in \Cref{fig:eu_gibbs}.}
  \label{fig:P_gibbs}
\end{figure*}

\Cref{fig:P_gibbs} illustrates the impact of coherency on the Gibbs free energy curves of the peritectic system at three temperatures relative to the stress-free peritectic temperature, $T_P$. In the stress-free limit ($\mathcal{W}=0$, left column), the system behaves classically. Unlike the eutectic case, where liquid--solid equilibria are both permissible (depending on the composition) at high temperatures, the peritectic system is defined by a competition between them. At temperatures below the peritectic point, the common tangent for $\beta+L$ is above the $\alpha+L$ and $\alpha+\beta$ lines. Introducing coherency stress ($\mathcal{W}>0$) changes equilibria in a manner distinct from the eutectic case due to the underlying arrangement of the free energy curves. In the peritectic system, the liquid curve is positioned to the side of the curves for the solids rather than between them as in the eutectic system. Consequently, when the coherent solid curve ($\alpha+\beta$) arches upward due to the strain energy, it does not intersect the liquid curve to form a broad three-phase region. This difference leads to conceptually distinct behavior: the lack of stabilization of the three-phase configuration by coherency stress observed in the eutectic system. Three-phase states that satisfy \Cref{eq:eqs} still exist in this system (red dashed curves in \Cref{fig:P_gibbs}(b,c)), but they correspond to saddle points (not minima) of the free energy (\Cref{sec:asymm} and Appendix~\ref{app:stability}), and their energies lie above those of the competing solid--liquid states.

\begin{figure*}[h]
  \centering
  \includegraphics[width=\linewidth]{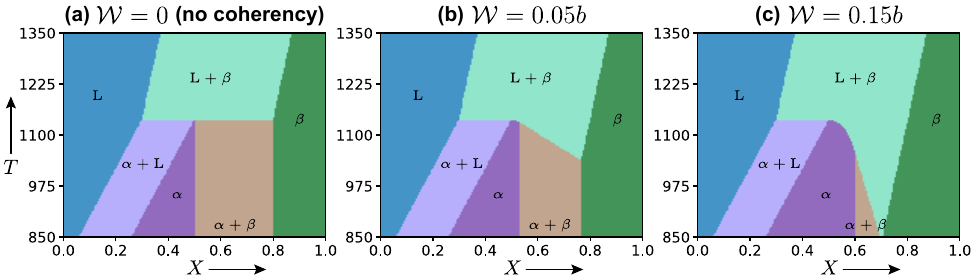}
  \caption{Phase equilibria maps for various coherency magnitudes, $\mathcal{W}$, expressed in terms of the reference free energy parameter of the solid phases ($b=b_\alpha=b_\beta$).}
  \label{fig:PD_PE}
\end{figure*}

Phase equilibria maps obtained for this peritectic system (\Cref{fig:PD_PE}) confirm that no three-phase configuration represents the global energy minimum across the full range of temperatures and alloy compositions consistent with the analysis at three selected temperatures in \Cref{fig:P_gibbs}. While coherency stress changes topology of the phase equilibria regions, there is no three-phase field like those observed in the eutectic system. Indeed, the coherency stress penalizes the two-solid equilibrium, whose region shrinks as the coherency stress increases. The $\alpha+\beta$ region is taken over by fields corresponding to single $\alpha$ and $\beta+L$ equilibria, whose energetics are not affected by the presence of the coherency stress between the two solid phases. For equal curvatures, the coherent $\alpha+\beta$ field closes at $\mathcal{W}=a(x^0_\beta-x^0_\alpha)^2=\SI{4.5}{\kilo\joule\per\mole}$ (Appendix~\ref{app:width}), above which $\alpha$ and $\beta$ do not coexist in equilibrium.

\subsection{Sensitivity to the free-energy parameters}
\label{sec:sensitivity}

The two case studies above use the symmetric parametrization of \Cref{tab:parameters}. To ensure that the findings and conceptual difference between the eutectic and peritectic systems generalize beyond this particular parameterization, we repeated the calculations for five modified parameter sets of each system, while keeping $b_L$ and $\varphi$ unchanged (\Cref{tab:sensitivity}): (i) unequal curvatures, $a_\alpha\ne a_\beta \ne a_L$; (ii) unequal reference energies of the solids, $b_\alpha \ne b_\beta$; (iii) both changes together; (iv) the reversed combination of curvatures and reference energies; and (v) unequal compositional offsets of the two solids from the liquid obtained by shifting $x^0_L$ from 0.5 to 0.4 in the eutectic system and from 0.3 to 0.2 in the peritectic system. For each set and for both $\mathcal{W}$ magnitudes, the equilibrium was determined over $0\le X\le 1$ and \SIrange{850}{1350}{\kelvin} with the reduced (polynomial) formulation described in \Cref{sec:comp} and Appendix \ref{app:mu}. The temperature extent, $\Delta T$, and the maximum composition width, $\Delta X$, of the three-phase field were resolved by following the three-phase states at fixed temperature and comparing their energies with those of all competing states.

Every parameter set for the eutectic system results in a three-phase field at both nonzero values of $\mathcal{W}$, whereas no stable three-phase state was found in any peritectic parameter set. Quantitative results are summarized in \Cref{tab:sensitivity}. Variations (i)--(iv) result in a change up to \SI{2.8}{\kelvin} for $\Delta T$ and 0.02 for $\Delta X$ from the baseline results. The largest change is from the unequal offsets in case (v), which is in agreement with an analytical derivation (Appendix~\ref{app:field}). The symmetric parametrization gives the smallest temperature extent for a given $\mathcal{W}/\varphi$ (\Cref{eq:dT}), however, its composition width is largest for symmetric spacing (Appendix~\ref{app:width}). These tests show that the fundamental difference in coherency effects on eutectic and peritectic systems generalizes to broader range of systems beyond equal curvatures, equal reference energies of the solids, or symmetric composition spacing.

\begin{table*}[h!]
\centering
\caption{Temperature extent, $\Delta T$, and maximum composition width, $\Delta X$, of the stable three-phase field in the eutectic system for the asymmetric parameter variations; energies are in \si{\kilo\joule\per\mole} and parameters not listed are those of \Cref{tab:parameters}.} 
\label{tab:sensitivity}
\begin{tabular}{llcccc}
\toprule
 & & \multicolumn{2}{c}{$\Delta T$ (K)} & \multicolumn{2}{c}{$\Delta X$} \\
\cmidrule(lr){3-4}\cmidrule(lr){5-6}
Set & Modification & $\mathcal{W}=1$ & $\mathcal{W}=3$ & $\mathcal{W}=1$ & $\mathcal{W}=3$ \\
\midrule
Baseline & -- & 29.9 & 89.8 & 0.567 & 0.500 \\
(i) & $a_\alpha,a_\beta,a_L=45,50,55$ & 29.9 & 89.8 & 0.565 & 0.494 \\
(ii) & $b_\alpha,b_\beta=20,22$ & 29.9 & 89.8 & 0.567 & 0.500 \\
(iii) & (i) and (ii) & 30.3 & 90.3 & 0.552 & 0.482 \\
(iv) & $a_\alpha,a_\beta,a_L=55,50,45$; & \multirow{2}{*}{30.6} & \multirow{2}{*}{92.6} & \multirow{2}{*}{0.555} & \multirow{2}{*}{0.492} \\
 & $b_\alpha,b_\beta=22,20$ & & & & \\
(v) & $x^0_L=0.4$ & 39.9 & 119.8 & 0.375 & 0.325 \\
\bottomrule
\end{tabular}
\end{table*}

\section{Discussion}
\label{sec:discussion}

\subsection{Difference in coherency effects in eutectic and peritectic systems}
\label{sec:asymm}

The central result of this study obtained with the proposed model (\Cref{sec:model}) is that coherency stress affects eutectic and peritectic invariant equilibria in conceptually different ways. In eutectic systems, the classical invariant point expands into a finite three-phase ($\alpha+\beta+L$) equilibrium field. In peritectic systems, coherency stress progressively destabilizes the two-solid ($\alpha+\beta$) equilibrium and can entirely suppress the peritectic equilibrium. This contrast has not been previously identified and is the main outcome of the present work extending the Cahn--Larché framework \cite{cahn1984simple} to three-phase equilibria involving a liquid.

Violation of the Gibbs phase rule and breakdown of the common-tangent construction in the presence of coherency stress observed in this study is consistent with prior findings in solid-state systems \cite{cahn1984simple,johnson1987phase,chiang1989coherent}.  The $z_\alpha z_\beta \mathcal{W}$ term introduces explicit dependence of the system free energy on phase fractions, which is absent in the classical thermodynamics that treats the free energy as a function of composition and temperature only (at constant pressure). Our new finding is that the third phase need not itself carry elastic energy for the topology of the equilibrium to change. Although an incoherent liquid does not contribute to elastic energy $g_e$, the classical eutectic point still expands into a finite three-phase field, whereas, in the peritectic systems, the same penalty suppresses the two-solid equilibrium instead. The deciding factor between the two effects is the composition of the liquid relative to those of the coherent solid pair, as discussed below.

The effects on phase equilibria in a eutectic system reported here are consistent with the findings of Roytburd and Sung \cite{roytburd1993formation}, whose model shares with the present work a dependence of the two-phase free energy on phase fractions through a fraction-dependent interaction term. However, in their treatment, the elastic energy arises from direct and substrate-mediated epitaxial strain and can be either positive or negative, which determines the change in the phase diagram. In the present work, the $\mathcal{W}$ term originates from the lattice misfit between bulk solid phases without external constraints and is  thus strictly non-negative. As a result, the coherency effect on the phase equilibria is controlled by the relationship between the liquid and solid compositions, which gives rise to the eutectic–peritectic asymmetry.

The fundamental difference in coherency effects on eutectic and peritectic systems follows from the ways the phase fractions of a three-phase mixture can change under fixed phase compositions. In a eutectic system, the liquid composition lies between those of the two solids ($x_\alpha < x_L < x_\beta$), and the exchange is $L\rightleftharpoons\alpha+\beta$: liquid is replaced by a mixture of two solids so that $z_\alpha$ and $z_\beta$ increase or decrease together. When coherency stress raises the free energy of the $\alpha+\beta$ mixture, the system lowers its total energy by replacing part of the strained solid with stress-free liquid. The liquid thus acts as a buffer that dilutes the coherency penalty of the solid pair and stabilizes three-phase states over a range of temperatures and compositions below the stress-free eutectic temperature. In a peritectic system, the liquid composition lies outside the range spanned by the two solids (e.g., $x_L < x_\alpha < x_\beta$) and the exchange is $L+\beta\rightleftharpoons\alpha$. The liquid cannot substitute for the solid--solid mixture, and one solid forms at the expense of the other. Coherency then destabilizes the three-phase states and penalizes the two-solid field, which is progressively taken over by the single-phase and solid--liquid fields that carry no coherency energy.

This interpretation can be formalized. A three-phase stationary solution of \Cref{eq:eqs} is a minimum of the total free energy and a candidate equilibrium state only if the liquid composition lies strictly between those of the two solids by a margin that grows with $\mathcal{W}$. This follows from the condition (derived in Appendix~\ref{app:stability}):

\begin{equation}
\left(x_\alpha - x_L\right)\left(x_\beta - x_L\right) < -\frac{\mathcal{W}S}{2} < 0,
\label{eq:sign_condition}
\end{equation}

\noindent where $S = \sum_i z_i/2a_i$ is the compositional susceptibility of the three-phase mixture. The same condition holds for any chemical free energies that are convex at the phase compositions, with $S=\sum_i z_i/g_i''(x_i)$ (Appendix~\ref{app:stability}), so it does not depend on the quadratic form of \Cref{eq:gs,eq:gL}. The condition is local: whether a candidate state is the equilibrium is determined from comparison with all competing states, as in the phase equilibria maps. The three-phase field does not appear in the peritectic systems because \Cref{eq:sign_condition} is violated even though three-phase stationary solutions of \Cref{eq:eqs} exist. Violation of this condition means those solutions are saddle points (and hence not minima) of the free energy at any $(X,T)$.

The condition in \Cref{eq:sign_condition} can be violated  even in a eutectic system by a sufficiently large $\mathcal{W}$. For equal curvatures, this happens at $\mathcal{W}=4a|\delta_\alpha\delta_\beta|$ with $\delta_i=x^0_i-x^0_L$ (Appendix~\ref{app:stability}), i.e., at \SI{18}{\kilo\joule\per\mole} ($\approx0.86b$) for the eutectic system in \Cref{tab:parameters}, well above the values considered here. Coherency can thus destroy the three-phase field it otherwise causes.

Note that, for systems with dissimilar energy curvatures, the condition is strictly in terms of the equilibrium compositions $x_i$ and \textit{not} the reference compositions $x^0_i$. The two compositional sequences of phases are identical when the three curvatures are equal ($a_\alpha = a_\beta=a_L$), in which case $x_i - x_j = x^0_i - x^0_j$. When the curvatures differ, the criterion must strictly use the equilibrium compositions, $x_i$.

\subsection{Limitations}
\label{sec:lims}

This section discusses the consequences of the simplifications employed in the present model.

First, the Gibbs free energy functions are represented as parabolas (\Cref{eq:gs,eq:gL}) centered on the reference composition of each phase. This approximation, used as the second-order expansion of any smooth molar Gibbs energy about its minimum, omits the configurational entropy of mixing, $RT[x\ln x+(1-x)\ln(1-x)]$, whose composition derivative diverges near the pure component limits. The omission removes these singularities and can therefore displace the phase boundaries at dilute compositions. However, the stability condition in \Cref{eq:sign_condition} is not specific to a particular functional form: it holds for any chemical free energies that are convex at the phase compositions (Appendix~\ref{app:stability}). In addition, the tests in \Cref{sec:sensitivity} show that the eutectic--peritectic contrast persists under variations of the quadratic parameters. Phase boundary positions will nonetheless quantitatively shift when realistic thermodynamic functions are used (e.g., from CALPHAD databases \cite{lukas2007computational}).

Second, both solid phases are assumed to have identical, isotropic elastic constants. This assumption, inherited from the original Cahn--Larché model \cite{cahn1984simple}, simplifies the elastic energy to the Vegard-type expression used here (\Cref{eq:g_e}) but neglects elastic inhomogeneity between phases, which can introduce additional compositional and morphological dependencies \cite{johnson1987phase,voorhees2004thermodynamics}. More importantly, \Cref{eq:g_e} assumes the coherency strains of the solid pair are relaxed by the liquid in proportion to its amount. The three-phase field predicted here is therefore conditional on this relaxation mechanism and the specific form of $g_e$ adopted in \Cref{eq:g_e}. Application to real alloys requires quantifying the $g_e$ dependence on fractions and phase morphologies as well as extension of the model to elastic inhomogeneity and anisotropy (e.g., leveraging recent developments \cite{spatschek2016scale,phan2019coherent,phan2019modeling}).

Third, the equilibrium maps presented here represent global free energy minima at each $(X, T)$ point. Johnson and Voorhees \cite{johnson1987phase,johnson1988coherent} showed that coherent two-phase systems can admit multiple linearly stable equilibrium states under identical thermodynamic conditions, unlike the unique common-tangent equilibrium of incoherent systems. Experimentally, which equilibrium state is realized depends on the kinetic pathway: solidification rate, nucleation sequence, and thermal history -- all influence whether the system reaches the global minimum or is trapped in a metastable configuration. The maps in \Cref{fig:eutectic-pd,fig:PD_PE} should therefore be interpreted as equilibrium limits.  

\subsection{Relevance to real systems and outlook}
\label{sec:real}

To assess the magnitude of the predicted effects in real materials, we applied the model to the Cu--Ag eutectic system and put our results in the context of coherent elastic energies calculated from elastic properties or estimated in literature. The elastic energy is discussed in terms of the dimensionless ratio, $\mathcal{W}/b$, and summarized in \Cref{tab:materials}.

\begin{figure*}[h]
  \centering
  \includegraphics[width=\linewidth]{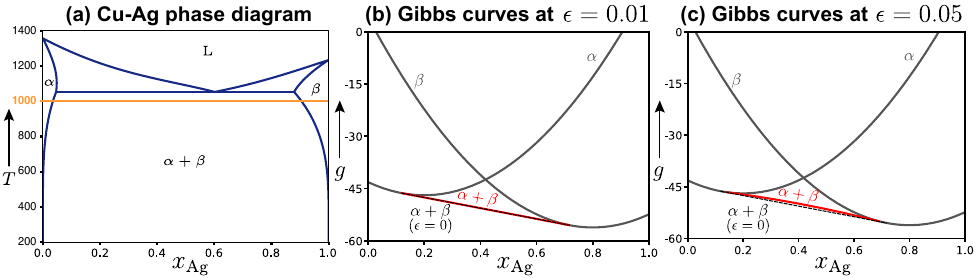}
  \caption{Assessment of predicted effects for a realistic eutectic system exemplified by Cu--Ag: (a) phase diagram; parabolic approximations of the Gibbs free energies with deviations from the common tangent corresponding to (b) a realistic effective lattice misfit of \SI{1}{\percent} ($\mathcal{W}=\SI{0.12}{\kilo\joule\per\mole}$) and (c) an upper-end value of \SI{5}{\percent} ($\mathcal{W}=\SI{2.96}{\kilo\joule\per\mole}$). The common tangent for $\alpha+\beta$ is shown as a dashed line in (b,c). Temperature in kelvin, Gibbs energy in \SI{}{\kilo\joule\per\mole}.}
  \label{fig:Cu-Ag}
\end{figure*}

For  Cu--Ag, we extracted the Gibbs free energy minima for the Cu-rich and Ag-rich solid solutions at \SI{1000}{\kelvin} from Thermo-Calc and matched our parabolic free energy functions to these minima. The elastic strain energy was evaluated using the elastic properties of copper ($E=\SI{110}{\giga\pascal}$, $\nu=0.34$ \cite{callister2020materials}), its molar volume, $V=\SI{7.1e-6}{\cubic\metre\per\mole}$, and a variable lattice misfit between the equilibrium Cu-rich and Ag-rich phases (\Cref{fig:Cu-Ag}). For a realistic effective coherent misfit of \SI{1}{\percent} representative of the strain ($\epsilon\approx0.01$) that can be sustained elastically without  misfit dislocations in typical metallic systems, the resulting $\mathcal{W}$ is small relative to the characteristic chemical free-energy differences (\Cref{fig:Cu-Ag}(b)). Even when pushing the misfit to a higher-end value ($\epsilon = 0.05$) of the small-strain assumptions underlying the Cahn--Larché formulation, the elastic contribution to the free energy remains modest compared to the chemical components, and the deviation of the coherent free energy from the classical common tangent for the $\alpha+\beta$ mixture is small accordingly (\Cref{fig:Cu-Ag}(c)). Quantitatively, \Cref{eq:g_e} gives $\mathcal{W}=\SI{0.12}{\kilo\joule\per\mole}$ at $\epsilon=0.01$ and $\mathcal{W}=\SI{2.96}{\kilo\joule\per\mole}$ at $\epsilon=0.05$, i.e., $\mathcal{W}/b=0.006$ and $0.14$ (\Cref{tab:materials}). The elastic energy at a coherently sustainable misfit of \SI{1}{\percent} is thus about an order of magnitude below the smaller value used in \Cref{sec:results}, whereas the value at $\epsilon=0.05$ nearly coincides with the larger one ($\mathcal{W}=\SI{3}{\kilo\joule\per\mole}\approx0.15b$). For an order-of-magnitude estimate of the corresponding three-phase field, we identify $\varphi$ with the mean entropy of fusion of Cu and Ag, $\Delta S_f=\SI{9.46}{\joule\per\mole\per\kelvin}$ \cite{dinsdale1991sgte}, and use the symmetric equal-curvature result $\Delta T=\mathcal{W}/(2\varphi)$ (Appendix~\ref{app:field}). This gives $\Delta T\approx\SI{6}{\kelvin}$ at $\epsilon=0.01$ and $\approx\SI{160}{\kelvin}$ at $\epsilon=0.05$, which indicate that the broadening of the $L+\alpha+\beta$ field is minor at a misfit that Cu--Ag can sustain coherently and becomes substantial only if a misfit of several percent remains coherent.

Based on these estimates, the coherency effect in a system like Cu–Ag is rather  limited by microstructure  than by thermodynamics. The equilibrium misfit between the Cu-rich and Ag-rich solid solution phases is close to \SI{10}{\percent}, which is well beyond sustainable coherency. Misfit dislocations nucleate above a critical particle size that falls steeply with increasing $\epsilon$ \cite{matthews1974defects,porter2009phase}, which is why the two solid phases are incoherent in conventionally processed bulk Cu--Ag. The misfits of 1 and \SI{5}{\percent} considered above emulate effective residual misfits of partially coherent microstructures. The large-$\mathcal{W}$ regime could thus be accessible in nanoscale precipitates, thin films and rapidly solidified microstructures.

Coherent solid phases also form directly from the melt in the $L\to\gamma+\gamma'$ eutectic of Ni-base superalloys, which involves a lattice misfit of up to about \SI{0.5}{\percent} \cite{pollock2006nickel}. Microstructural observations indicate that this transformation occurs over a temperature range  \cite{d2016role,pollock2006nickel}. With the elastic properties of nickel ($E=\SI{207}{\giga\pascal}$, $\nu=0.31$ \cite{callister2020materials}, $V=\SI{6.59e-6}{\cubic\metre\per\mole}$), \Cref{eq:g_e} gives $\mathcal{W}\approx\SI{0.05}{\kilo\joule\per\mole}$ at $\epsilon=0.005$, and the scaling $\Delta T\approx\mathcal{W}/(2\Delta S_f)$ with $\Delta S_f\approx\SI{10}{\joule\per\mole\per\kelvin}$ for Ni \cite{dinsdale1991sgte} gives only about \SI{2}{\kelvin}. The coherency contribution to the temperature range of this reaction is therefore expected to be small compared with those of multicomponent thermodynamics and kinetics.

The scaling $\mathcal{W} \propto E \epsilon^2$ indicates that substantially larger effects are expected in systems with high elastic stiffness and/or large sustainable coherent misfit. Candidate systems include refractory alloys based on Mo or W, whose Young's moduli of \SIrange{330}{410}{\giga\pascal} are about three times that of copper. With the properties of tungsten ($E=\SI{407}{\giga\pascal}$, $\nu=0.28$ \cite{callister2020materials}, $V=\SI{9.53e-6}{\cubic\metre\per\mole}$), the same effective misfits of 1 and \SI{5}{\percent} give $\mathcal{W}=\SI{0.54}{\kilo\joule\per\mole}$ and \SI{13.5}{\kilo\joule\per\mole} ($\mathcal{W}/b=0.026$ and 0.64), i.e., about 4.6 times the Cu--Ag values, so that an effective misfit of about \SI{2.4}{\percent} would already give the larger value used in \Cref{sec:results}. Tungsten is also nearly isotropic elastically (Zener ratio \num{1.01} \cite{featherson1963elastic}) satisfying the isotropy assumption in \Cref{eq:g_e}.

\begin{table*}[h!]
\centering
\caption{Coherency energy parameter, $\mathcal{W}$, for selected systems compared with the values used in \Cref{sec:results} and with a chemical energy scale of each system (in \si{\kilo\joule\per\mole}). $\mathcal{W}/b$ ratio uses the fixed reference $b=\SI{21}{\kilo\joule\per\mole}$. Energies are per mole of atoms for the metallic systems and per mole of formula units for the pseudobinary compounds. The misfits for the metallic systems are assumed effective values; details of the estimates are given in Appendix~\ref{app:tab2}. Footnotes: \textsuperscript{a}$b_L-b$; \textsuperscript{b}enthalpy of fusion \cite{dinsdale1991sgte}, averaged over Cu and Ag; \textsuperscript{c}leading interaction coefficient of the regular- or subregular-solution model.}
\label{tab:materials}
\small
\begin{tabular}{lcccc}
\toprule
System & $\epsilon$ & $\mathcal{W}$ (\si{\kilo\joule\per\mole}) & $\mathcal{W}/b$ & Chem.\ scale \\
\midrule
Model systems & -- & 1; 3 & 0.048; 0.14 & 19\textsuperscript{a} \\ 
Cu--Ag (Eq.~\ref{eq:g_e}, Cu) & 0.01 & 0.12 & 0.006 & 12.3\textsuperscript{b} \\ 
Cu--Ag (Eq.~\ref{eq:g_e}, Cu) & 0.05 & 2.96 & 0.14 & 12.3\textsuperscript{b} \\ 
Ni-base $\gamma/\gamma'$ (Eq.~\ref{eq:g_e}) & 0.005 & 0.05 & 0.002 & 17.5\textsuperscript{b} \\ 
Refractory (Eq.~\ref{eq:g_e}, W)  & 0.01 & 0.54 & 0.026 & 52.3\textsuperscript{b} \\ 
Refractory (Eq.~\ref{eq:g_e}, W) & 0.05 & 13.5 & 0.64 & 52.3\textsuperscript{b} \\ 
LiFePO$_4$--FePO$_4$  \cite{cogswell2012coherency} & anisotropic & [2.7,4.2] & [0.13,0.20] & 11.1\textsuperscript{c} \\ 
Mg$_2$Si--Mg$_2$Sn \cite{yi2018strain} & [0.062,0.065] & [17,19] & [0.8,0.9] & 17.9\textsuperscript{c} \\ 
PbTe--PbS \cite{doak2012coherent} & 0.084 & 20.7 & 0.99 & 27.9\textsuperscript{c} \\
\bottomrule
\end{tabular}
% \smallskip
\end{table*}

Larger misfits still are available in ordered intermetallics with large lattice parameter differences between disordered and ordered phases as well as strain-engineered semiconductor alloys, such as Ge--Sn \cite{JUNG2025121369}. However, as in Cu--Ag, the attainable effect is largely limited by the retention of coherency.

In thermoelectric PbTe--PbS alloys, coherency strain qualitatively alters the miscibility gap and stabilizes metastable solid solutions that decompose into coherent nanoscale precipitates  upon annealing \cite{girard2013analysis}. In Mg$_2$Si–Mg$_2$Sn, coherent strain energy has been shown to entirely suppress the chemical spinodal and phase decomposition at the processing temperatures \cite{yi2018strain}. For these systems, the coherency energy can be compared directly with the chemical interaction that drives phase separation. For PbTe--PbS with the lattice parameter difference of \SI{8.4}{\percent}, Doak and Wolverton \cite{doak2012coherent} calculated the leading coefficients of subregular-solution fits to the coherent and incoherent mixing enthalpies as \num{74.62} and \SI{289.5}{\milli\electronvolt} per cation, respectively; their difference is the leading coefficient of the coherency strain energy, $\mathcal{W} = \SI{20.7}{\kilo\joule\per\mole}$, or $\mathcal{W}/b = 0.99$. This coherency energy offsets \SI{74}{\percent} of the incoherent interaction coefficient (\SI{27.9}{\kilo\joule\per\mole}) and lowers the maximum of the miscibility gap from \SI{1770}{\kelvin} for incoherent phase separation to a coherent spinodal maximum of \SI{610}{\kelvin}. Mg$_2$Si--Mg$_2$Sn reaches a comparable magnitude: a misfit of \SIrange{6.2}{6.5}{\percent} with the elastic properties of the two end members compiled by Yi et al.\ \cite{yi2018strain} gives $\mathcal{W}\approx\SIrange{16.5}{19}{\kilo\joule\per\mole}$ per mole of formula units ($\mathcal{W}/b\approx0.8$--0.9), comparable to the leading interaction parameter of their chemical model, \SI{17.9}{\kilo\joule\per\mole}. While these examples do not directly confirm the existence of a wide three-phase field, they demonstrate that the coherency strain energy can be significant on a scale comparable to the chemical energy.

For a final supporting example, we cite LiFePO$_4$ as a system for which the coherency energy has been quantified and tested against experiments. Cogswell and Bazant \cite{cogswell2012coherency} calculated the elastic energy coefficient of a coherent phase boundary from first-principles elastic constants and the measured lattice mismatch as \SI{0.19}{\giga\pascal} for the elastically preferred $\{101\}$ orientation, which  corresponds to $\mathcal{W} = \SIrange{2.7}{4.2}{\kilo\joule\per\mole}$, or $\mathcal{W}/b = 0.13$ to $0.20$ based on the molar volume of LiFePO$_4$ (Appendix~\ref{app:tab2}). Their fitted regular-solution parameter is $\Omega = \SI{11.1}{\kilo\joule\per\mole}$, so the coherency energy amounts to \SIrange{24}{38}{\percent} of $\Omega$. This calculated coherency energy serves as an independent support for the elastic energy range explored in this study.

The present framework could benefit from extensions that would bring it closer to experimental relevance. Incorporating realistic  free energy functions from CALPHAD databases \cite{lukas2007computational} combined with a justified elastic energy for the solid--liquid morphology would enable quantitative predictions for specific alloy systems. Marschall et al.\ \cite{marschall2025incorporating} have recently demonstrated such integration for solid-state coherent equilibria in multi-component alloys, which could serve as a viable pathway for extending the present framework to realistic systems. Allowing for partial coherency through a composition- or size-dependent loss-of-coherency criterion would bridge the gap between the fully coherent limit treated here and the incoherent limit assumed in conventional phase diagram calculations. Coupling the modified equilibrium conditions to a solidification model (e.g., Scheil-type calculation \cite{porter2009phase}) would predict how the altered equilibria affect freezing paths, microsegregation, mushy zones, and microstructural length scales in casting and additive manufacturing \cite{debroy2018additive}. Finally, phase-field simulations that resolve the elastic fields in the coherent solids could test the fraction dependence of $g_e$ assumed in \Cref{eq:g_e} for specific solid--liquid morphologies and could capture morphological evolution during solidification through coherent three-phase regions, including the competition between equilibrium and metastable states discussed in \Cref{sec:lims}.

A more immediately testable extension is to invariant reactions among three solid phases. Li$_x$FePO$_4$ is a natural candidate as it undergoes a eutectoid reaction involving a disordered phase at elevated temperature \cite{delacourt2005existence,dodd2006phase}, and its coherency energy is already known from the calculations discussed above \cite{cogswell2012coherency}. For a eutectoid reaction, the elastic energy may depend on all three phase fractions because all phases are solid and every pair-wise interface may be coherent. The present developments  serve as a foundation for such generalizations.

\section{Conclusions}

We extended the Cahn--Larché thermodynamic framework to binary alloys in which two coherent solid phases coexist with an incoherent liquid that relaxes their coherent energy and applied it to eutectic- and peritectic-forming systems. The key findings are as follows. 

\begin{enumerate}

\item Coherency stress transforms the classical eutectic invariant point into a finite three-phase ($L+\alpha+\beta$) equilibrium field spanning a continuous range of compositions and temperatures. The temperature extent of this field increases with the coherency energy (linearly for equal curvatures), while its width along composition decreases. For free energy curves with equal curvatures, the temperature extent, $\Delta T$, can be obtained in closed form, which reduces to $\mathcal{W}/(2\varphi)$ in the simplest case. As the liquidus is unaffected by coherency, the three-phase field forms by depression of the solidus by up to $\Delta T$.

\item In systems whose equilibrium compositions are peritectically ordered, coherency stress does not stabilize a three-phase field. Instead, it progressively destabilizes the two-solid ($\alpha+\beta$) equilibrium, which is consumed by single-phase and solid–liquid fields and can be suppressed entirely at sufficiently large strain energies.

\item The difference in the behavior between eutectic and peritectic systems has a geometric and analytical interpretation. Geometrically, when the liquid composition lies between those of the two solids (eutectic), the liquid can act as a thermodynamic buffer against the coherency penalty; when it lies outside (peritectic), it cannot. Analytically, the criterion for three-phase equilibrium  takes the form $(x_\alpha - x_L)(x_\beta - x_L) < -\mathcal{W}S/2$, which is a necessary condition for the three-phase state to be a nondegenerate free-energy minimum and holds for any chemical free energy that is convex at the phase compositions.

\item The eutectic--peritectic difference persists when the curvatures, the reference energies of the solids, and the composition spacing are varied from the symmetric parametrization. 

\item Estimates of the coherency energy for real materials vary in a wide range from  $\le\SI{0.1}{\kilo\joule\per\mole}$ in Ni-base $\gamma/\gamma'$ and Cu--Ag systems at coherently sustainable misfits (well below present magnitudes) to about \SI{20}{\kilo\joule\per\mole} in PbTe--PbS and Mg$_2$Si--Mg$_2$Sn systems (well above present magnitudes). For Cu--Ag, the corresponding three-phase field would span a few kelvin at a misfit of \SI{1}{\percent} but more than \SI{100}{\kelvin} if a misfit of \SI{5}{\percent} was coherent.

\end{enumerate}

These results show that coherency stress between solid phases can fundamentally alter three-phase equilibria involving a liquid in binary alloys when the liquid relaxes the coherency strains of the solid pair, and suggest that such effects may be significant in systems with large coherent misfits and high elastic stiffness, such as refractory alloys, ordered intermetallics, and strain-engineered semiconductors.

\section*{Code availability}

The code for this paper is available on GitHub: \\ 
\url{https://github.com/materials-informatics-az/coherency}.

\section*{Declaration of competing interest}

The authors declare that they have no known competing financial interests or personal relationships that could have appeared to influence the work reported in this paper.

% ==============================================================================
% TRANSITION TO SUPPLEMENTARY MATERIAL
% ==============================================================================

\clearpage % Start on a new page

% 1. Reset figure counter & update representation (e.g., S1, S2)
\setcounter{figure}{0}
\renewcommand{\thefigure}{S\arabic{figure}}

% 2. Reset subsection counter & update representation (e.g., A1, A2)
\setcounter{subsection}{0}
\renewcommand{\thesubsection}{A\arabic{subsection}}

% 3. Reset equation counter & update representation (e.g., (A1), (A2))
\setcounter{equation}{0}
\renewcommand{\theequation}{A\arabic{equation}}

% 4. Update cleveref label formats
\crefname{figure}{Figure}{Figures}
\crefname{subsection}{Appendix}{Appendices}
\crefname{equation}{Equation}{Equations} % Optional: \Cref{eq:label} -> "Equation A1"

% (Optional) Reset and prepend table numbering as well if needed:
% \setcounter{table}{0}
% \renewcommand{\thetable}{S\arabic{table}}
% ==============================================================================

\section*{Appendix: Analytical solution for parabolic free energies}

\subsection{Reduced two-variable formulation}
\label{app:reduced}

Here is a solution alternative to \Cref{sec:model}. First, we fix the phase fractions $z_\alpha$, $z_\beta$, and hence $z_L = 1 -z_\alpha - z_\beta$. We then use the same method of Lagrange multipliers as in \Cref{sec:model} for minimizing the total energy over phase compositions $x_\alpha$, $x_\beta$, and $x_L$ only. In this case, $\mathcal{W}z_\alpha z_\beta$ and $\sum z_ib_i$ are both constant terms and the only constraint is the mass balance (\Cref{eq:constr}a). In this case, the Lagrangian is
\begin{equation}
\mathcal{L} = \sum_i z_i a_i (x_i - x_i^0)^2 - \lambda\left(\sum_i z_i x_i - X\right).
\end{equation}

\noindent Its derivative in respect to the composition of the $i^\text{th}$  phase is
\begin{equation}
\frac{\partial \mathcal{L}}{\partial x_i} = 2 z_i a_i (x_i - x_i^0) - \lambda z_i = 0,
\end{equation}

\noindent which allows us to express the $i^\text{th}$ composition without $z_i$:
\begin{equation}
x_i = x_i^0 + \frac{\lambda}{2 a_i}.
\label{eq:app_comp}
\end{equation}

\noindent Substitution of this composition expression into the mass balance
\begin{equation}
\sum_i z_i\left(x_i^0 + \frac{\lambda}{2a_i}\right) = X
\end{equation}

\noindent allows us to introduce two quantities: the composition offset, $M$, of the alloy from the fraction-weighted reference composition, $\sum z_ix_i^0$, and the compositional susceptibility of the mixture, $S$, expressed as
\begin{subequations}
\begin{align}
M &= X - \sum_i z_i x_i^0, \\
S &= \sum_i \frac{z_i}{2 a_i}.
\end{align}
\end{subequations}

\noindent Then the total energy in \Cref{eq:g_total} can be re-written with only two unknowns:
\begin{equation}
G(z_\alpha, z_\beta) = \frac{M^2}{2S} + \sum_i z_i b_i - \varphi T z_L + \mathcal{W} z_\alpha z_\beta,
\label{eq:g_total_2}
\end{equation}

\noindent recalling that $z_L = 1 -z_\alpha - z_\beta$. With this approach, the feasible set is a triangle and the minimum of the $G$ function in the triangle gives the phase(s) in equilibrium. The minimum at a corner means only one phase, an edge -- two phases, and interior -- all three phases. This solution assumes that the compositions of the present phases lie within $[0,1]$. 

\subsection{The diffusion potential and closed-form solution}
\label{app:mu}

The Lagrange multiplier introduced above has a physical meaning. Comparing \Cref{eq:app_comp} with the derivative of \Cref{eq:gs}, $\mathrm{d}g_i/\mathrm{d}x_i = 2a_i(x_i - x_i^0)$, gives
\begin{equation}
\lambda = \frac{\mathrm{d}g_i}{\mathrm{d}x_i} = \mu
\label{eq:lambda_is_mu}
\end{equation}

\noindent for every phase simultaneously so that $\lambda$ is the diffusion potential $\mu=\mu_B-\mu_A$ of Larché and Cahn \cite{larche1973linear,larche1985overview}, and the first two equalities of \Cref{eq:eqs} are satisfied identically in this formulation. Since $\mu$ fixes all three phase compositions at once, it is the natural coordinate in which to solve the remaining equilibrium conditions. Combining \Cref{eq:app_comp} with \Cref{eq:lambda_is_mu} expresses composition
\begin{equation}
x_i = x_i^0 + \frac{\mu}{2a_i},
\end{equation}

\noindent whose substitution into \Cref{eq:gs,eq:gL} gives
\begin{subequations}
\begin{align}
g_i &= \frac{\mu^2}{4a_i} + b_i \text{ for } i \in {\alpha,\beta} \text{ and} \\
g_L &= \frac{\mu^2}{4a_L} + b_L- {\varphi}T,
\end{align}
\label{eq:mu_param}
\end{subequations}

\noindent so a single scalar determines all three phase compositions and all three phase energies. It is convenient to introduce the following for solid phases $i \in \{\alpha, \beta\}$
\begin{equation}
\begin{split}
\delta_i &= x^0_i - x^0_L, \\
\qquad k_i &= \frac{1}{a_i} - \frac{1}{a_L}, \\
\qquad B_i &= b_i - b_L + \varphi T,
\end{split}
\label{eq:appdefs}
\end{equation}

\noindent where $\delta_i$ measures the compositional offset of solid $i$ from the liquid and $k_i$ the difference in compositional susceptibility. With these, the last two equalities of \Cref{eq:eqs} reduce to explicit quadratic functions in $\mu$,
\begin{subequations}
\label{eq:zofmu}
\begin{align}
\mathcal{W} z_\alpha &= \frac{k_\beta}{4}\mu^2 + \delta_\beta \mu - B_\beta, \\
\mathcal{W} z_\beta  &= \frac{k_\alpha}{4}\mu^2 + \delta_\alpha \mu - B_\alpha.
\end{align}
\end{subequations}

Combining \Cref{eq:zofmu} with the mass balance leaves a single polynomial equation in $\mu$, of degree at most three for every combination of phases, which can be solved in a closed form. When the three curvatures are equal (which is the case for the symmetric parameterization in \Cref{tab:parameters}), $k_\alpha = k_\beta = 0$ and the polynomial reduces to a linear equation.

\subsection{Temperature extent of the three-phase field}
\label{app:field}

A three-phase state requires $z_\alpha > 0$, $z_\beta > 0$ and $z_L = 1 - z_\alpha - z_\beta > 0$. Written with \Cref{eq:zofmu} as bounds on $\varphi T$, and introducing $F_i(\mu)$, for solid phases $i \in\{\alpha,\beta\}$
\begin{equation}
F_i(\mu) = \delta_i \mu + \frac{k_i}{4}\mu^2 - \left(b_i - b_L\right),
\label{eq:Fdef}
\end{equation}

\noindent these three conditions become boundaries
\begin{subequations}
\label{eq:three_bounds}
\begin{align}
\varphi T &< F_\alpha(\mu), \\
\varphi T &< F_\beta(\mu), \\
\varphi T &> \frac{F_\alpha(\mu) + F_\beta(\mu) - \mathcal{W}}{2}.
\end{align}
\end{subequations}

\noindent The third boundary is the arithmetic mean of the first two displaced by $\mathcal{W}/2$.

\Cref{eq:three_bounds} has the following consequences. First, at $\mathcal{W} = 0$, the three-phase region collapses to the classical invariant point because the three boundaries become concurrent where $F_\alpha = F_\beta$. It is therefore the coherency energy alone that opens three-phase equilibrium into a field. Second, the upper vertex of the field lies where $F_\alpha = F_\beta$ -- a condition without $\mathcal{W}$, which means the apex of the stationary field is pinned at the stress-free invariant temperature. For the stable eutectic fields computed in this work, this apex is the top of the field, which therefore extends downward as $\mathcal{W}$ increases. For $b_\alpha = b_\beta$ and equal curvatures, this gives $T_E = (b_L - b_\alpha)/\varphi = \SI{1137.7}{\kelvin}$ for the parameters in \Cref{tab:parameters}. Third, writing $\Sigma(\mu) = F_\alpha(\mu) - F_\beta(\mu)$, the field is non-empty where $|\Sigma(\mu)| < \mathcal{W}$. The three vertices lie at $\Sigma = 0$, $-\mathcal{W}$ and $+\mathcal{W}$, corresponding to $(z_\alpha, z_\beta, z_L) = (0,0,1)$, $(1,0,0)$ and $(0,1,0)$.

For equal curvatures, $F_i$ are straight lines, $\Sigma$ is linear with slope $D = \delta_\alpha - \delta_\beta$, and the field is a triangle whose vertices are separated in $\mu$ by $\mathcal{W}/|D|$. Evaluating $F_\alpha$ and $F_\beta$ at those vertices gives the vertical displacement of each below the apex as $\mathcal{W}\delta_\alpha/D$ and $-\mathcal{W}\delta_\beta/D$. In a eutectic system with $\delta_\alpha < 0 < \delta_\beta$, both displacements are positive and the height of the triangle is the larger of the two:
\begin{equation}
\Delta T = \frac{\mathcal{W}}{\varphi}\,\frac{\max\left(|\delta_\alpha|, |\delta_\beta|\right)}{|\delta_\alpha| + |\delta_\beta|},
\label{eq:dT}
\end{equation}

\noindent which results in
\begin{equation}
\frac{\mathcal{W}}{2\varphi} \le \Delta T < \frac{\mathcal{W}}{\varphi}.
\end{equation}

\noindent The equality on the left corresponds to $|\delta_\alpha| = |\delta_\beta|$. Neither $a$ nor $b$ enters the equation: the curvature cancelled in \Cref{eq:zofmu} when $k_i = 0$, and the reference energies enter only through the location of the apex. For the eutectic parameters of \Cref{tab:parameters}, the compositions are symmetric, so $\Delta T = \mathcal{W}/2\varphi$, giving 
\SI{29.9}{\kelvin} at $\mathcal{W}=\SI{1}{\kilo\joule\per\mole}$ and \SI{89.8}{\kelvin} at $\mathcal{W}=\SI{3}{\kilo\joule\per\mole}$. The symmetric parametrization thus realizes the lower bound of \Cref{eq:dT} and produces the smallest 
temperature extent for a given $\mathcal{W}/\varphi$. The composition width, in contrast, is largest for symmetric spacing (\Cref{app:width}).

In a peritectic system, $\delta_\alpha$ and $\delta_\beta$ have the same sign, the two displacements have opposite signs, and the stationary field extends both above and below the stress-free peritectic temperature with total extent $(\mathcal{W}/\varphi)(\delta_\alpha + \delta_\beta)/|\delta_\alpha - \delta_\beta|$. That this field is nevertheless never observed is shown in \Cref{app:stability}.

Finally, the two solid--liquid boundaries contain no $\mathcal{W}$. Indeed, $\mathcal{W}z_\alpha z_\beta$ vanishes identically when either solid fraction is zero, so the $\alpha$--$L$ and $\beta$--$L$ equilibria are the classical common tangents at every coherency energy. The liquidus is therefore unchanged. The solidus (i.e., the lower boundary of liquid-containing states) is depressed by $\Delta T$ at most -- for alloys whose composition lies within the base of the three-phase field. For example, for the eutectic in \Cref{tab:parameters} at $\mathcal{W}=\SI{3}{\kilo\joule\per\mole}$,  the depression is \SI{89.8}{\kelvin} for $0.25\le X\le0.75$ but \SI{18}{\kelvin} at $X=0.21$.

\subsection{Composition width of the three-phase field}
\label{app:width}

For equal curvatures, the alloy composition is a function of $\mu$ and $\varphi T$:
\begin{equation}
\begin{split}
X = x^0_L + \mu\left[\frac{1}{2a} + \frac{2\delta_\alpha \delta_\beta}{\mathcal{W}}\right] - \\
- \frac{\delta_\alpha\left(b_\beta - b_L\right) + \delta_\beta\left(b_\alpha - b_L\right)}{\mathcal{W}} - \\
- \frac{\varphi T \left(\delta_\alpha + \delta_\beta\right)}{\mathcal{W}},
\end{split}
\label{eq:Xofmu}
\end{equation}

\noindent so the three-phase field is a triangle in the $(X,T)$ plane as well. Its two lower vertices are at $\delta_\alpha - \mathcal{W}/(2aD)$ and $\delta_\beta + \mathcal{W}/(2aD)$ from the apex. The edge joining them therefore has length
\begin{equation}
\Delta x^0 - \frac{\mathcal{W}}{a\,\Delta x^0},
\label{eq:base_edge}
\end{equation}

\noindent with $\Delta x^0 = x^0_\beta - x^0_\alpha$. The composition projection of this edge (\Cref{eq:base_edge}) is also the composition width of the coherent $\alpha+\beta$ field: the three-phase triangle rests directly on top of the two-solid field so that the base of the triangle is equal to the width of the field. The maximum horizontal width of the triangle is measured at the height of its intermediate vertex:
\begin{equation}
\begin{split}
\Delta X = 
2\min\left(|\delta_\alpha|,|\delta_\beta|\right) - \\ 
- \frac{\mathcal{W}}{2a\max\left(|\delta_\alpha|,|\delta_\beta|\right)}.
\end{split}
\label{eq:dX}
\end{equation}

\noindent \Cref{eq:base_edge,eq:dX} coincide when $|\delta_\alpha| = |\delta_\beta|$, which is the case for the parameters of \Cref{tab:parameters}. Unlike $\Delta T$, both expressions depend on the curvature and both decrease linearly with $\mathcal{W}$: the field grows in temperature and shrinks in composition as coherency increases. Setting \Cref{eq:base_edge} to zero gives $\mathcal{W}_c = a(\Delta x^0)^2$ -- the coherency energy at which the coherent two-solid field closes. For coherent two-phase equilibrium, $\mathcal{W}_c$ is the Williams point \cite{williams1984calculation,cahn1984simple}. The three-phase minimum disappears earlier, at $4a|\delta_\alpha\delta_\beta|\le\mathcal{W}_c$, where $\Delta X$ in \Cref{eq:dX} reaches zero (\Cref{app:stability}). The two thresholds coincide only for symmetric spacing, $|\delta_\alpha|=|\delta_\beta|$, as for the eutectic in \Cref{tab:parameters}, for which both  equal $W_c =\SI{18}{\kilo\joule\per\mole}$.

\subsection{Second-order condition and the eutectic--peritectic criterion}
\label{app:stability}

The solutions described above identify the stationary states. Whether a stationary state is a minimum can be determined by the second derivatives of \Cref{eq:g_total_2} organized into a Hessian:
\begin{equation}
\mathbf{H} = \frac{1}{S}\,\mathbf{v}\mathbf{v}^{\mathsf{T}} + \mathcal{W}\begin{pmatrix} 0 & 1 \\ 1 & 0\end{pmatrix}
\label{eq:hessian}
\end{equation}
\noindent with $\mathbf{v} = \left(x_\alpha - x_L,\; x_\beta - x_L\right)$. The determinant of the Hessian is
\begin{equation}
\begin{split}
\det \mathbf{H} &= \\
& -\mathcal{W}\left[\frac{2\left(x_\alpha - x_L\right)\left(x_\beta - x_L\right)}{S} + \mathcal{W}\right],
\end{split}
\end{equation}

\noindent which leads to \Cref{eq:sign_condition}: the liquid composition must lie strictly between those of the two solids, by a margin that grows with $\mathcal{W}$. For $\mathcal{W}>0$, $\mathbf{H}$ is positive definite, and the stationary state is a nondegenerate local minimum if \Cref{eq:sign_condition} holds. A negative determinant implies a saddle point, whereas equality is a degenerate case.

This single condition accounts for the central result of the present work. In a eutectic system $x_\alpha < x_L < x_\beta$, the product is negative, and a three-phase minimum is possible provided $\mathcal{W}$ is not too large. In a peritectic system $x_L < x_\alpha < x_\beta$, the product is positive, $\det \mathbf{H} < 0$ for every  $\mathcal{W}>0$ and any convex chemical free energies, and the stationary state of \Cref{app:field} is a saddle point instead of a minimum. The peritectic three-phase solution therefore exists as a solution of \Cref{eq:eqs} but cannot be realized as a three-phase equilibrium -- the same conclusion as numerical results (\Cref{fig:PD_PE} and Section~\ref{sec:sensitivity}) and geometric interpretation.

The criterion is stated in terms of the equilibrium compositions $x_i$, not the reference compositions $x^0_i$. From \Cref{eq:mu_param}, the two orderings coincide when the curvatures are equal but may differ otherwise, so the ordering of the $x^0_i$ is a reliable proxy only within the equal-curvature parametrization used in \Cref{sec:results}. For equal curvatures, $S=1/(2a)$ and the condition becomes $\mathcal{W} < 4a|\delta_\alpha||\delta_\beta|$, which is the coherency energy at which $\Delta X$ of \Cref{eq:dX} reaches zero closing the three-phase field in composition. point.

This expression does not rely on the quadratic form of the $g_i$. For any chemical free energies that are convex at the phase compositions, $\partial G/\partial z_\alpha=g_\alpha-g_L-\mu(x_\alpha-x_L)+\mathcal{W}z_\beta$, which similarly applies to $z_\beta$. Setting these derivatives to zero recovers the last two equalities of \Cref{eq:eqs}. Differentiating again, the chemical energy contributes only through the change of the diffusion potential with the fractions, $\partial\mu/\partial z_j=-(x_j-x_L)/S$, which yields the same $\mathbf{H}$ with $S=\sum_i z_i/g_i''(x_i)$; for \Cref{eq:gs,eq:gL}, $g_i''=2a_i$.

\subsection{Details on coherency estimates for example systems}
\label{app:tab2}

Here we provide details of calculations for estimates in \Cref{tab:materials}. For the metallic systems, \Cref{eq:g_e} was evaluated with room-temperature Young's moduli and Poisson's ratios of polycrystalline Cu, Ni, and W \cite{callister2020materials} and the molar volumes of the pure metals, $V=7.1$, 6.59, and \SI{9.53e-6}{\cubic\metre\per\mole}, respectively. The misfits of these systems are assumed effective values (Section~\ref{sec:real}). Since elastic moduli decrease with temperature, room-temperature values tend to overestimate $\mathcal{W}$ near the solidus. The chemical scale listed for these systems is the enthalpy of fusion \cite{dinsdale1991sgte}, which corresponds to $b_L-b$ in the model systems.

For the pseudobinary systems, the misfit $\epsilon$ is based on the mean lattice parameter \cite{williams1984calculation}:
\begin{equation}
\epsilon = \frac{2 (a_\beta-a_\alpha)}{a_\beta+a_\alpha}
\label{eq:ref-misfit}
\end{equation}

For the PbTe--PbS system, the coherency energy was estimated directly from the first-principles calculations \cite{doak2012coherent} as the difference between the leading coefficients of the subregular-solution fits to the incoherent and coherent mixing enthalpies (Table I in \cite{doak2012coherent}). For Mg$_2$Si--Mg$_2$Sn, \Cref{eq:g_e} was evaluated with the Young's moduli, Poisson's ratios, and lattice parameters compiled by Yi et al.\ \cite{yi2018strain}, averaged over the two end members, and the mean molar volume per mole of formula units, $V\approx\SI{4.25e-5}{\cubic\metre\per\mole}$. The range reflects the lattice parameters between room temperature and the processing temperature and isotropic moduli obtained either directly from their table or as Voigt--Reuss--Hill averages of their single-crystal constants. The leading interaction parameter in the Mg$_2$Si--Mg$_2$Sn system, $\Omega = 3 \times \SI{5.95}{\kilo\joule\per\mole}$ is per mole of formula units, which is converted from the originally reported normalization per mole of atoms  \cite{yi2018strain} (one of three sublattices mixes). The smaller subregular term of their model is neglected in this comparison. For the LiFePO$_4$ system, the lower and upper values correspond to composition differences between the coexisting phases of $\Delta x=0.8$ and 1, respectively. In this system, the lattice mismatch is strongly anisotropic so the coherency energy, $\mathcal{W}$, was obtained from the calculated strain-energy coefficient $B = \SI{0.19}{\giga\pascal}$ and the expression \cite{cogswell2012coherency}:
\begin{equation}
\mathcal{W} = \frac{1}{2}BV_m\left ( \Delta x \right)^2
\label{eq:W_from_B}
\end{equation}

\noindent with $V_m=\SI{4.38e-5}{\cubic\metre\per\mole}$ for LiFePO$_4$. \Cref{eq:W_from_B} is the coefficient of $z_\alpha z_\beta$ for two coexisting phases whose compositions differ by $\Delta x$.

\bibliography{refs}
\end{document}